\documentstyle[12pt]{article}
\newcommand{\be}{\begin{equation}}
\newcommand{\ee}{\end{equation}}
\newcommand{\bea}{\begin{eqnarray}}
\newcommand{\eea}{\end{eqnarray}}
\newcommand{\sirc}[1]{\stackrel{\circ}{#1}}
\begin{document}
\begin{flushright}
ITP-SB-99-19\\
hep-th/9905075\\
\end{flushright}

\begin{center}
{\LARGE Consistent 
nonlinear KK reduction of 11d supergravity on $AdS_7\times S_4$ and 
self-duality in odd dimensions}
\vskip1truecm

{\large\bf Horatiu Nastase,
 Diana Vaman and  
Peter van Nieuwenhuizen}\footnote{Research supported by NSF Grant 9722101;  
E-mail addresses: 
hnastase@insti.physics.sunysb.edu,\protect\\
dvaman@insti.physics.sunysb.edu,
vannieu@insti.physics.sunsyb.edu
} \\ {\large
Institute for Theoretical Physics,\\ S.U.N.Y. Stony Brook, NY 11794-3840,USA\\}
\vskip1truecm
\end{center}
\abstract{
\parbox {4.75 in}{

~~~~~We show that there exists a consistent 
truncation of 11 dimensional
supergravity to the 'massless' fields of maximal (N=4) 
7 dimensional gauged supergravity. 
We find the complete expressions for the nonlinear embedding of 
the 7 dimensional 
fields into the 11 dimensional fields, and check them by reproducing the d=7 
susy transformation laws from the d=11 laws in various sectors. 
In particular we determine 
explicitly the matrix U which connects the Killing spinors to the gravitinos 
in the KK ansatz, and the dependence of  
the 4-index field strength on the scalars.
This is the first time a complete nonlinear KK reduction of the original 
d=11 supergravity on a 
nontrivial compact space has been explicitly given.
 We need a first order
formulation for the 3 index tensor field $A_{\Lambda\Pi\Sigma}$ in d=11
 to reproduce the 7 dimensional result. 
The concept of 'self-duality in odd dimensions' is thus shown to originate 
from first order formalism in higher dimensions. For the AdS-CFT 
correspondence, our results imply that one can use 7d 
gauged supergravity (without further massive modes)
 to compute certain correlators in the d=6 (0,2) CFT at 
leading order in N. This eliminates an ambiguity in the formulation of the
correspondence.}}
\section*{}

~~~~~The question whether in general a consistent Kaluza-Klein (KK) truncation
 exists at the nonlinear level is an old problem. For tori, the 
consistency is easy to prove, but for more complicated compact spaces 
little is known.  
In supergravity (sugra), the truncation of d=11 sugra on $AdS_4\times S_7$ 
to maximal $d=4$ gauged sugra was intensively studied 15 years ago \cite{ck},  
culminating in a series of papers by de Wit and Nicolai 
\cite{dwn84,dwn86}. The interest in those days was to find realistic 4 dimensional 
models from
 spontaneous compactification of maximal 11 dimensional sugra.
Recent developments in the AdS-CFT correspondence 
\cite{mald,gkp,witten,3point,4point} have renewed interest in AdS 
compactifications to 5 and 7 dimensions. 
A crucial question is whether the $AdS_5\times S_5$ and $AdS_7\times S_4$
compactifications allow consistent truncations to the massless sector, 
because if
there does not exist a consistent truncation, massive fields have to be 
considered for computations of correlators in the corresponding CFT 
\cite{witten}.

de Wit and Nicolai first studied the KK reduction of 
the original formulation of d=11 sugra \cite{dwn84}, but then they found it
advantageous to construct a different formulation of d=11 sugra with a
local SU(8) invariance \cite{dwn86}. Because the local SU(8) invariance 
rotates Bianchi identities into field equations, the action for the SU(8)
sugra did not follow directly from  the 
original d=11 action \cite{dwn86}). For this SU(8) formulation 
the complete nonlinear KK reduction on $AdS_7\times S_4$ was given:
they  proposed nonlinear ans\"{a}tze and
checked the consistency of the KK truncation on $AdS_4\times S_7$ 
 as far as the bosonic and fermionic transformation rules are 
concerned. It may be that the SU(8) theory will turn out to be 
important for future research in string theory, but we prefer to work with 
the original formulation.
The connection was formulated in terms of a matrix U for which they derived 
an equation, but they only could solve this equation in certain 
sectors. (ref. \cite{dwn84}a, eq. (3.12) and ref. \cite{dwn84}b, eq. (2.14)).

In this article we analyze the KK reduction of d=11 sugra on 
 $AdS_7\times S_4$. This will allow us to go further 
than the work on $AdS_4\times S_7$. As we already mentioned, 
in ref. \cite{dwn84}
 d=11 sugra was first reformulated
 in a form with a local $SU(8)\times SO(1,3)$ in tangent space 
instead of the usual SO(1,10).
In our case, we choose not to go through the intermediate step of finding 
a formulation of 11d sugra with a corresponding local SO(5) invariance, 
hence we will directly 
work with the d=11 sugra as it is usually formulated.
In this letter we present our main results leaving the details of 
the calculations for a future publication \cite{wip}.

One of the reasons to study the $AdS_7\times S_4$ case 
is that we would like to understand the origin of the mysterious 
'self-duality in odd dimensions' \cite{tpn}
which appears in various supergravities in odd dimensions. 
To obtain the action for selfdual tensors one begins with the 
antisymmetric tensor from d=11 whose action is quadratic in derivatives, 
and introduces the square of an extra auxiliary
antisymmetric tensor field in the lower dimension
 by hand, rotates both tensors, factorizes the 
second order field equations into two field
equations linear in derivatives, and drops one of the factors.  
The end product is an action of the form $\epsilon FA+A^2$ \cite{tpn} 
which is dual to Chern-Simons theory for the abelian case \cite{dj} 
but not for the nonabelian case \cite{klr}.
In sugra the nonabelian version appears. 
\vspace{.5cm}
\\
$
\begin{array}{|lr|}
\hline
{\cal L}=-\frac{1}{2k^2}ER(E,\Omega )-\frac{E}{2}\bar{\Psi}_{\Lambda}\Gamma
^{\Lambda\Pi\Sigma}D_{\Pi}(\frac{\Omega +\hat{\Omega}}{2})\Psi_{\Sigma}
+\frac{E}{48}({\cal F}_{\Lambda\Pi\Sigma\Omega}{\cal F}^{\Lambda\Pi
\Sigma\Omega}-&\\
48{\cal F}^{\Lambda\Pi\Sigma\Omega}\partial_{\Lambda}A_{\Pi\Sigma\Omega})
-\frac{\sqrt{2}}{6}k\epsilon^{\Lambda_0...\Lambda_{10}}\partial_{\Lambda_0}
A_{\Lambda_1\Lambda_2\ \Lambda_3}\partial_{\Lambda_4}A_{\Lambda_5\Lambda_6
\Lambda_7}A_{\Lambda_8\Lambda_9\Lambda_{10}}&\\
-\frac{\sqrt{2}k}{16}E[\bar{\Psi}_{\Pi}\Gamma^{\Pi\Lambda_1...\Lambda_4\Sigma}
\Psi_{\Sigma}+12\bar{\Psi}^{\Lambda_1}\Gamma^{\Lambda_2\Lambda_3}\Psi^{
\Lambda_4}][\frac{1}{24}(F+\hat{F})_{\Lambda_1...\Lambda_4}] & (I.1)\\
\hline
\hline
\delta E_{\Lambda}^M=\frac{k}{2}\bar{\varepsilon}\Gamma^M\Psi_{\Lambda}
\;{\rm with}\; \Gamma^a=\tau^a\otimes\gamma_5;\Gamma^m=I\otimes\gamma^m
& (I.2)\\
\delta A_{\Lambda_1\Lambda_2\Lambda_3}=-\frac{\sqrt{2}}{8}\bar{\varepsilon}
\Gamma_{[\Lambda_1\Lambda_2}\Psi_{\Lambda_3]} &(I.3)\\
\delta \Psi_{\Lambda}=\frac{1}{k} D_{\Lambda}(\hat{\Omega})\varepsilon
+\frac{\sqrt{2}}{12}(\Gamma^{\Lambda_1...\Lambda_4}\;_{\Lambda}-8\delta
_{\Lambda}^{\Lambda_1}\Gamma^{\Lambda_2\Lambda_3\Lambda_4})
\varepsilon(\frac{1}{24}\hat{F}_{\Lambda_1...\Lambda_4})&\\
+ \frac{1}{24}(b\Gamma_{\Lambda}\;^{
\Lambda_1...\Lambda_4}\frac{1}{\sqrt{E}}B_{\Lambda_1...\Lambda_4}\nonumber
-a\Gamma
^{\Lambda_1\Lambda_2\Lambda_3}\frac{1}{\sqrt{E}}B_{\Lambda\Lambda_1\Lambda_2
\Lambda_3})\varepsilon\ &(I.4)\\
\delta 
B_{MNPQ}= \sqrt{E}\bar{\varepsilon}
(a\Gamma_{MNP}E_{Q}^{\Lambda}
R_{\Lambda}(\Psi )+b \Gamma_{MNPQ\Lambda}R^{\Lambda}
(\Psi )) &(I.5)\\
\hline
\end{array}
$
\\
$$ Table\; I$$
\\
Maximal  (N=4) sugra in 
7 dimensions has a 3 index tensor $S_{\alpha\beta\gamma ,A}$ (A=1,5)
with a self-dual action. Because in the linearized KK reduction 
one needs to introduce
an auxiliary field $B_{\alpha\beta\gamma}\sim {}^*B_{\alpha\beta\gamma\delta}$ 
 in d=7 to construct this 
action \cite{pnt},
we will start from
a first 
order formulation for $A_{\Lambda\Pi\Sigma}$ in d=11, and deduce 
self-duality.
The 11 dimensional Lagrangian we start from and its 
supersymmetry transformations rules are given in the Table I, where 
$\Lambda, \Pi,...=0,10$
are curved vector indices and M,N,...=0,10 are flat vector indices.
The action  
is invariant under the susy laws 
 for any value of the real constants $a$ and $b$. We will fix them later
by the requirement of consistent truncation.

The field ${\cal F}_{\Lambda\Pi\Sigma\Omega}$ is an independent field. 
$\hat{F}_{\Lambda\Pi\Sigma\Omega}$ denotes the usual curl 
$F_{\Lambda\Pi\Sigma\Omega}= 24\partial_{[\Lambda}A_{\Pi\Sigma
\Omega]}$ plus the $\Psi\Gamma\Gamma\Psi$-terms which make it
supercovariant.
\footnote{By replacing the term ${\cal F}F$ in (I.1) by ${\cal F}\hat{F}$,
the terms $(\bar{\Psi}\Gamma\Gamma \Psi)(F+\hat{F})$ get absorbed. 
Then the ${\cal
F}$ field equation reads ${\cal F}=\hat{F}$ and becomes supercovariant. We have
not been able to absorb the remaining four-fermi terms by using our new
first order formulation.} Similarly $\hat{\Omega}_{\Lambda}\;^{MN}$ is the 
usual supercovariant  spin connection. \cite{peter} 
Finally, $R_{\Lambda}(\Psi)$ is the gravitino field equation, 
$R_{\Lambda}(\Psi)=\frac{1}{E}\frac{\delta{\cal L}}{\delta
\bar{\Psi}_{\Lambda}}$. 
It is convenient to redefine  ${\cal F}_{\Lambda\Pi\Sigma\Omega}$ 
by introducing an auxiliary tensor density $B_{MNPQ}$:
\be
{\cal F}_{\Lambda\Pi\Sigma\Omega}=\partial_{\Lambda}A_{\Pi\Sigma\Omega}+23
\; terms+B_{MNPQ}E^{-1/2}E_{\Lambda}^M...E_{\Omega}^Q
\ee

The action in table I contains a new first-order formulation as far as 
the 3-index tensor field $A_{\Lambda\Pi\Sigma}$ is concerned. There exists
also a first order formulation of d=11 sugra with an independent spin
connection $\Omega$ \cite{cfgpp}. 
Initially we tried to deduce the d=7 selfduality 
from this field $\Omega$, but this did not work \cite{wip}. Instead 
we will work with a second order formalism for the spin connection.
\vspace{.5cm}
\\
$
\begin{array}{|lr|}
\hline
e^{-1}{\cal L}=-\frac{1}{2}R+\frac{1}{4}m^2(T^2-2T_{ij}T^{ij})-\frac{1}{2}
P_{\alpha ij}P^{\alpha ij}-\frac{1}{4}({\Pi_A}^i{\Pi_B}^j 
F_{\alpha\beta}^{AB})^2
&\\
+\frac{1}{2}({{\Pi^{-1}}_i}^AS_{\alpha\beta\gamma ,A})^2 
+\frac{1}{48}me^{-1}\epsilon^{\alpha\beta\gamma\delta\epsilon\eta\zeta}
\delta^{AB}S_{\alpha\beta\gamma ,A}F_{\delta\epsilon\eta\zeta ,B}-
\frac{1}{2}\bar{\psi}_{\alpha}\tau^{\alpha\beta\gamma}\bigtriangledown_{\beta}
\psi_{\gamma}-
&\\
\frac{1}{2}\bar{\lambda}^i\tau^{\alpha}\bigtriangledown_{\alpha}
\lambda_i
-\frac{1}{8} m(8T^{ij}-T\delta^{ij})\bar{\lambda}_i\lambda_j+\frac{1}{2}
mT^{ij}\bar{\lambda}_i\gamma_j\tau^{\alpha}\psi_{\alpha}+\frac{1}{2}\bar{\psi}
_{\alpha}\tau^{\beta}\tau^{\alpha}\gamma^i\lambda^jP_{\beta ij}
&\\
+\frac{1}{8}mT\bar{\psi}_{\alpha}\tau^{\alpha\beta}\psi_{\beta}
+\frac{1}{16}\bar{\psi}_{\alpha}(\tau^{\alpha\beta\gamma\delta}-2\delta^{\alpha
\beta}\delta^{\gamma\delta})\gamma_{ij}\psi_{\delta}{\Pi_A}^i{\Pi_B}^j
F_{\beta\gamma}^{AB}+
&\\
\frac{1}{4}\bar{\psi}_{\alpha}\tau^{\beta\gamma}
\tau^{\alpha}\gamma_i\lambda_j{\Pi_A}^i{\Pi_B}^jF_{\beta\gamma}^{AB}
+\frac{1}{32}\bar{\lambda}_i\gamma^j\gamma_{kl}\gamma^i\tau^{\alpha\beta}
\lambda_j{\Pi_A}^k{\Pi_B}^lF_{\alpha\beta}^{AB}+\frac{im}{8\sqrt{3}}
\bar{\psi}_{\alpha}(\tau^{\alpha\beta\gamma\delta\epsilon}+
&\\
6\delta^
{\alpha\beta}\tau^{\gamma}\delta^{\delta\epsilon})\gamma^i\psi_{\epsilon}
{{\Pi^{-1}}_i}^A S_{\beta\gamma\delta ,A}
-\frac{im}{4\sqrt{3}}\bar{\psi}_{\alpha}(\tau^{\alpha\beta\gamma\delta}-3
\delta^{\alpha\beta}\tau^{\gamma\delta})\lambda^i{{\Pi^{-1}}_i}^A S_{\beta
\gamma\delta ,A}-
&\\
\frac{im}{8\sqrt{3}}\bar{\lambda}^i\tau^{\alpha\beta\gamma}
\gamma^j\lambda_i{{\Pi^{-1}}_i}^A S_{\beta\gamma\delta ,A}
-i\frac{e^{-1}}{16\sqrt{3}}\epsilon^{\alpha\beta\gamma\delta\epsilon\eta\zeta}
\epsilon_{ABCDE}\delta^{AG}S_{\alpha\beta\gamma, G}F_{\delta\epsilon}^{BC}
F_{\eta\zeta}^{DE}
&\\
+\frac{m^{-1}}{8}e^{-1}\Omega_5[B]-
\frac{m^{-1}}{16}e^{-1}\Omega_3[B]& (II.1)\\
\hline
\hline
\delta e_{\alpha}^a =\frac{1}{2}\bar{\epsilon}\tau^a\psi_{\alpha}
& (II.2)\\
{\Pi_A}^i{\Pi_B}^j\delta B_{\alpha}^{AB}=\frac{1}{4}\bar{\epsilon}
\gamma^{ij}\psi_{\alpha}+\frac{1}{8}\bar{\epsilon}\tau_{\alpha}\gamma^k
\gamma^{ij}\lambda_k
&(II.3)\\
\delta S_{\alpha\beta\gamma ,A}=-\frac{i\sqrt{3}}{8m}{\Pi_A}^i(2
\bar{\epsilon}\gamma_{ijk}\psi_{[\alpha}+\bar{\epsilon}\tau_{[\alpha}\gamma^l
\gamma_{ijk}\lambda_l){\Pi_B}^j{\Pi_C}^kF_{\beta\gamma]}^{BC}
&\\
-\frac{i\sqrt{3}}{4m}\delta_{ij}{\Pi_A}^j D_{[\alpha}
(2\bar{\epsilon}\tau_{\beta}\gamma^i\psi_{\gamma]}+\bar{\epsilon}\tau_{\beta
\gamma ]}\lambda^i ) 
+\frac{i\sqrt{3}}{12}\delta_{AB}{{\Pi^{-1}
}_i}^B(3\bar{\epsilon}\tau_{[\alpha\beta}\gamma^i\psi_{\gamma]}-\bar{\epsilon}
\tau_{\alpha\beta\gamma}\lambda^i ) &(II.4)\\
{{\Pi^{-1}}_i}^A\delta {\Pi_A}^j=\frac{1}{4}(\bar{\epsilon}\gamma_i
\lambda^j+\bar{\epsilon}\gamma^j\lambda_i)& (II.5)\\
\delta\psi_{\alpha}=\bigtriangledown_{\alpha}\epsilon-\frac{1}{20}mT
\tau_{\alpha}\epsilon -\frac{1}{40}(\tau_{\alpha}\;^{\beta\gamma}-8
\delta_{\alpha}^{\beta}\tau^{\gamma})\gamma_{ij}\epsilon {\Pi_A}^i{\Pi_B}^j
F_{\beta\gamma}^{AB}
&\\
+\frac{im}{10\sqrt{3}}(\tau_{\alpha}\;^{\beta\gamma\delta}
-\frac{9}{2}\delta_{\alpha}^{\beta}\tau^{\gamma\delta})\gamma^i\epsilon 
{{\Pi^{-1}}_i}^AS_{\beta\gamma\delta ,A} & (II.6)\\
\delta\lambda_i=\frac{1}{16}\tau^{\alpha\beta}(\gamma_{kl}\gamma_i-\frac{1}{5
}\gamma_i\gamma_{kl})\epsilon {\Pi_A}^k{\Pi_B}^l F_{\alpha\beta}^{AB}
+\frac{im}{20\sqrt{3}}\tau^{\alpha\beta\gamma}(\gamma_i\; ^j-4\delta_i^j)
\epsilon{{\Pi^{-1}}_j}^A S_{\alpha\beta\gamma ,A}
&\\
+\frac{1}{2}m(T_{ij}-\frac{1}{5}T)\gamma^j\epsilon+\frac{1}{2}\tau^{\alpha}
\gamma^j \epsilon P_{\alpha ij}&(II.7)\\
\hline
\end{array}
$
\\
$$Table\;II$$
\\
We expand d=11 sugra
around the $AdS_7\times S_4$ background given by $F_{\mu\nu\rho\sigma}
=3/\sqrt{2}m\epsilon_{\mu\nu\rho\sigma} $ $det(\sirc e_\mu^m)$ 
where $m^{-1}$ is the radius of $S_4$ and $\sirc e_\mu^m(x)$ is the background
vielbein ($\mu,\nu,...=1,4 $ are
curved indices and m,n,... are flat indices).

Maximal gauged sugra in d=7 \cite{ppn} has the action and 
susy laws of Table II.
 Here
$\alpha,\beta,...=0,6$ are curved vector indices and a,b,...=0,6 are flat 
vector indices.
The Dirac matrices in d=7 and d=4 are denoted by $\tau^a$ and $\gamma^m$, 
respectively.
The model has a local $SO(5)_g$ gauge group for which A,B,... =1,5 are vector
indices, while I,J,... =1,4 are spinor indices.
The scalars ${\Pi_A}^i$ parametrize the coset $SL(5,{\bf R})/SO(5)_c$
but in the gauged model the  rigid $SL(5,{\bf R})$ symmetry of the 
action is lost and replaced by the $SO(5)_g$ gauge invariance.
The indices 
i,j,...=1,5 are $SO(5)_c$ vector indices and $I',J'$,...=1,4  are spinor 
indices.
The model has the folowing fields: the vielbein ${e_\alpha}^a$, the 
4 gravitinos $\psi^{I'}_\alpha$, the $SO(5)_g$ vector $B_\alpha^{AB}=-
B_\alpha^{BA}$, the scalars ${\Pi_A}^i$, the antisymmetric tensor 
$S_{\alpha\beta\gamma\,A}$ and the spin 1/2 fields $\lambda_i^{I'}$ 
(vector-spinors under $SO(5)_c$).They have the correct mass-terms which 
ensure 'masslessness' in d=7 AdS space \cite{mtn}.
In (II.1) $T_{ij}={{\Pi^{-1}}_i}^A{{\Pi^{-1}}_j}^B\delta_{AB}$, 
$\Omega_3[B]$ and  $\Omega_5[B]$ are the Chern-Simons forms for $B_{\alpha
}^{AB}$ (normalized to $d\Omega_3[B]=(TrF^2)^2$ and $d\Omega_5[B]=(TrF^4)$).
The tensor $P_{\alpha\,ij}$ and the connection $Q_{\alpha\,ij}$ (appearing in 
the covariant derivatives $\bigtriangledown_{\alpha}$)
 are the symmetric and antisymmetric parts of $
(\Pi^{-1})^A_i\left({\delta_A}^B \partial_\alpha + g B_{\alpha\,A}\;^B\right)
{\Pi_B}^k \delta_{kj}$, respectively. Here $D_{\alpha}$ has both a $Q_{\alpha
ij}$ and a $P_{\alpha ij}$ piece.

We begin the KK reduction with the usual ansatz for the 11d vielbein:
\be
{E_\Lambda}^M=
\pmatrix{
{e_\alpha}^a(y)\Delta^{-1/5} (y,x)& 
B_{\alpha}^{\mu}(y,x){E_\mu}^m(y,x)\cr
0& {E_\mu}^m (y,x)
}
\ee
\be
{E_M}^\Lambda =
\pmatrix{
{e_a}^\alpha \Delta^{1/5}& -B_{\alpha}^{\mu}(y,x){e_a}^\alpha \Delta^{1/5}\cr
0& {E_m}^\mu 
}
\; ; {E_m}^\mu {E_\mu}^n={\delta_m}^n\; ; 
{e_a}^\alpha {e_\alpha}^b={\delta_a}^b
\ee
where $B_{\alpha}^{\mu}(y,x)=-2 B_{\alpha}^{AB}(y)V^{\mu AB}(x)$ with 
$V^{\mu AB}$ Killing vectors on $S_4$ 
\footnote{One has $V_{\mu}^{AB}=
Y^{[A}\partial_{\mu}Y^{B]}$ with strength unity, where $Y^A=\frac{1}{4}
(\gamma^A)_{IJ}\bar{\eta}^I\gamma_5
\eta^J$ is real and satisfies $\sum_A(Y^A)^2=1$. We use 'modified' Majorana 
spinors, $\bar\eta^I=\eta^{I,T}C_{-}^{(4)}=(\eta^K)^{\dagger}\gamma_5\tilde
\Omega^{IK}$ with $(C_{-}^{(4)})_{..}$ numerically equal to $(\tilde{\Omega}
)_{..}$ and $\tilde{\Omega}^{..}$. 
The matrices
$\gamma^A$ and $\gamma^i$ are in both cases given by $\{i\gamma^m\gamma^5,
\gamma^5 \}$.
Furthermore, $\bar\varepsilon=\varepsilon^T C^{(11)}=
\varepsilon^\dagger i\Gamma^0$ with $C^{(11)}=C^{(7)} \otimes C^{(4)}_{-}$, 
so that $\bar\varepsilon (y,x)=\bar\epsilon_{I'} (y) {U^{I'}}_I (y,x) 
\bar\eta^I (x) \Delta^{-1/10}\sqrt{\gamma_5}$. }. 
The rescaling by $\Delta^{-1/5}$ where $\Delta\equiv
det {E_\mu}^m (y,x)/ \det{\sirc{e}_\mu}^m(x)$ brings the d=7 Einstein
action in canonical form.
 
We also redefine  the d=11 gravitino field $\Psi_\Lambda$ in terms of a 
field $\psi_\alpha (y,x)$ and a field $\psi_m(y,x)$ which lead to
 the canonical 
gravitino and Dirac Lagrangians in d=7
\be
{E_a}^\Lambda \Psi_\Lambda=
\Delta^{1/10} \gamma_5^{-1/2} {e_a}^\alpha \psi_\alpha-\frac{1}{5}
\tau_a \gamma_5 \gamma^m {E_m}^\Lambda \Psi_\Lambda\;\;\;;\;\;\;
{E_m}^\Lambda \Psi_\Lambda =\Delta^{1/10}\gamma_5^{1/2} \psi_m
\ee
We formulate the KK reduction of the fermions in terms of  $\psi_\alpha$ and
$\psi_m$ 
\bea
\psi_{\alpha}(y,x)&=&\Delta^{-1/10}(y,x)\gamma_5^{-1/2}e_{\alpha}^a(y)
(\Psi_a(y,x)+\frac{1}{5}\tau_a\gamma_5\gamma^m\Psi_m(y,x))\nonumber\\&=&
\psi_{\alpha I'}(y)U^{I'}\;_I(y,x)\eta^I(x)\\
\psi_m(y,x)&=&\Delta^{-1/10}(y,x)\gamma_5^{1/2}\Psi_m(y,x)\nonumber\\
&=&\lambda_{I'J'K'}(y)U^{I'}\;_I(y,x)U^{J'}\;_J(y,x)U^{K'}\;_K(y,x)
\eta_m^{IJK}(x)
\\
\epsilon (y,x)&=&\Delta^{1/10}(y,x)\gamma_5^{-1/2}\varepsilon(y,x)\;
{\rm with}\; \sqrt{\gamma_5}=\frac{1}{2}(1-i)(1+i\gamma_5)\nonumber\\
&=&\epsilon_{I'}(y)U^{I'}\;_I(y,x)\eta^I(x) \label{eps}
\eea
where $\eta^I$ are Killing spinors on $S_4$ (
$\sirc{D}_\mu \eta^I=\frac{im}{2}\gamma_\mu \eta^I $). We normalize them to
$\bar{\eta}^I\eta^J=\tilde{\Omega}^{IJ}$.
The expansion into spherical harmonics is the same as  in \cite{pnt}
except that we added a matrix ${U^{I'}}_I\; (y,x)$  which interpolates 
between $SO(5)_g$ and $SO(5)_c$. In ref \cite{dwn84}, a SU(8) matrix U was 
found to be needed to obtain consistency of the KK reduction in certain
sectors, but then a reformulation of the theory with 
full local SU(8) invariance was constructed \cite{dwn86}. 
We introduce the matrix U as in \cite
{dwn84}, but we shall not go to a different formulation 
of d=11 sugra.

 For consistency of our results for the 
transformation rules the matrix U needs to satisfy 
\be
U^{I'}\;_I\tilde{\Omega}^{IJ}{U^{J'}}_J=\tilde{\Omega}^{I'J'}
\rightarrow \tilde{\Omega}^{IJ}{U^{J'}}_J\Omega_{J'I'}=-{(U^{-1})^I}_{I'}
\label{uinv}
\ee
For example (II.2) follows from (I.2) only if (\ref{uinv}) holds.
Here $\Omega$ and $\tilde{\Omega}$ are the Usp(4) 
invariant tensors used to lower and raise the spinor indices, satisfying
$\tilde{\Omega}^{IJ}\Omega_{KJ}=\delta^I_K$ and $\tilde{\Omega}^{I'J'}
\Omega_{K'J'}=\delta^{I'}_{K'}$. Since $\Omega$ is
the charge conjugation matrix,
this restricts U to be an SO(5) matrix in the spinor representation.

The ansatz for the expansion of $E_{\mu}^m$ into spherical harmonics 
is found from the result in (II.3) that $
{\Pi_A}^i{\Pi_B}^j\delta B_{\alpha}^{AB}=\frac{1}{4}\bar{\epsilon}
\gamma^{ij}\psi_{\alpha}+\frac{1}{8}\bar{\epsilon}\tau_{\alpha}\gamma^k
\gamma^{ij}\lambda_k $. 
The first term in (II.3) gives the following result:
\bea
iE^\mu _m (U V^m U^T)^{I'J'}&=&-\Delta^{1/5}{(\Pi^{-1})
_i}^A{(\Pi^{-1})_j}^B
 V^{\mu}_{AB} (\gamma^{ij})^{I'J'}\label{vie}\\
E_m^{\mu}&=&i\frac{1}{4}\Delta^{1/5}{(\Pi^{-1})_i}^A{(\Pi^{-1})_j}^B
 V^{\mu}_{AB} Tr(\gamma^{ij} U V_m U^T\Omega)\label{viel}
\eea

By substituting $E^{\mu}_m$ back 
 into (\ref{vie}), we get a consistency condition
on the matrix U,
\be
\frac{1}{4}{(\Pi^{-1})_i}^A{(\Pi^{-1})
_j}^B V^{\mu}_{AB} Tr(\gamma^{ij} U V_m U^T\Omega)(U V^m U^T)^{I'J'}
={(\Pi^{-1})
_i}^A{(\Pi^{-1})_j}^B V^{\mu}_{AB} (\gamma^{ij})^{I'J'}\label{u1}
\ee
where the Killing vector $V^{m IJ}$ is given by $V^{m\;IJ}=iV^m_{AB}
(\gamma^{AB})^{IJ}=-i V^m_{AB}(\gamma^{AB}\tilde{\Omega})^{IJ}$.
We note that U=1 is not consistent, therefore we indeed need the 
matrix U. 
Using this ansatz, the second term in (II.3)
also matches the corresponding term on the left hand side, provided one 
identifies $\lambda^k_{I'}$ with $3i(\gamma^k)^{J'K'}\lambda_{I'J'K'}$. 

Then, by calculating $\Delta$ , we get
\be
\Delta^{-6/5}= {(\Pi^{-1})_i}^A  {(\Pi^{-1})_j}^B \delta^{ij} Y_A Y_B
\equiv T^{AB} Y_AY_B\label{del}
\ee
where $Y_A=\frac{1}{4}(\gamma_A)_{IJ}\bar{\eta}^I\gamma_5\eta^J$ is the 
basic scalar spherical harmonic on $S_4$. 
We can then extract $\delta {\Pi_A}^i$ from $\delta(\Delta^{-6/5})$ 
and comparison with (II.5) gives another condition on U:
\be
Y_A (U \gamma^A\tilde\Omega U^T)^{I'J'}= \Delta^{3/5}  {(\Pi^{-1})_i}^A 
(\gamma^{i})^{I'J'} Y_A \label{urel}
\ee
Equations (\ref{uinv}), 
(\ref{u1}) and (\ref{urel}) are all we need to know about U to 
prove all results on consistency.

At this point we have come in 7 dimensions as far as others in 4 dimensions. 
However, we have been able to find the  solution for U. First of all, we 
have been able to show that (\ref{u1}) follows from (\ref{urel}), so that 
(\ref{urel}) is the crucial equation. 
The covariant solution of (\ref{urel}) built out of $Y^A$ and $v_i$
 is unique and reads
\bea
U&=&-\sqrt{\frac{1+v_iY^i}{2}}+\frac{Y^Av_i\delta^{Bi}\gamma_{AB}}
{\sqrt{2(1+v_iY^i)}}
\nonumber\\
v_i&=&{{(\Pi^{-1})}_i}^A Y_A\Delta^{3/5}\label{usol}
\eea
It was determined by moving one of the U matrices in (\ref{urel}) to the 
right-hand side yielding $UY\llap/  =v\llap/  U$ 
and expanding the $4\times 4$ matrix U on the basis 
$1, \gamma_A ,\gamma_{AB}$. Covariance restricts U to $f_1+f_2Y\llap/ 
+f_3v\llap/  +f_4 Y^Av_j\delta^{Bj} \gamma_{AB}$, where $f_j$ depends only on
$Y\cdot v$.
Requiring (\ref{uinv}) and (\ref{urel}) leads
to (\ref{usol}).

Next we turn to the ansatz for $F_{\Lambda\Pi\Sigma\Omega}$. At the linearized
level it contains the fluctuations in $g_{\mu\alpha}$ and $g_{\mu\nu}$, and 
fluctuations in $A_{\alpha\beta\gamma}$ \cite{pnt}. At the nonlinear level,
an ansatz for $F_{\Lambda\Pi\Sigma\Omega}$ containing only the fluctuations
in $g_{\mu\alpha}$ and yielding the correct Chern-Simons actions in d=7 
was given in \cite{hmm,fhmm}. 
We now present the complete expression for 
$F_{\Lambda\Pi\Sigma\Omega}$; it contains the results of \cite{pnt} and 
\cite{hmm,fhmm} (at the limit when the scalar fields are set to zero) 
and it satisfies the Bianchi identities.
\bea
\frac{\sqrt{2}}{3m}
F_{\mu\nu\rho\sigma}
&=&\epsilon_{\mu\nu\rho\sigma}\sqrt{\det \sirc{g}}\left[1+
\frac{1}{3}\left(\frac{T}{Y_A Y_B T^{AB}}-5\right)\right.\nonumber\\&&\left.
-\frac{2}{3} \left(\frac{Y_A (T^2)^{AB} Y_B}{(Y_A T^{AB} Y_B)^2}-1\right)
\right]
\label{mnpr}
\\
\frac{\sqrt{2}}{3m}
F_{\mu\nu\rho\alpha}&=&
\partial_{[\mu} \left(\epsilon_{ABCDE}B_\alpha^{AB}
C_\nu^C C_{\rho ]}^D \frac{T^{EF} Y_F}{Y\cdot T\cdot Y}\right)\nonumber\\
&&+\sqrt{\sirc{g}}\epsilon_{\mu\nu\rho\sigma}C^\sigma_A
\frac{1}{3}\left(\frac{\partial_\alpha T^{AB} Y_B}{Y_A T^{AB} Y_B}-
\frac{T^{AB} Y_B}{(Y_A T^{AB} Y_B)^2} (Y_C \partial_\alpha T^{CD} Y_D)
\right)
\label{mnra}\\
\frac{\sqrt{2}}{3m}
F_{\mu\nu\alpha\beta}&=&\frac{2}{3}\left[\partial_{[\alpha}
\left(\epsilon_{ABCDE}
B_{\beta ]}^{AB} C_\mu^C C_\nu^D \frac{T^{EF} Y_F}{Y\cdot T\cdot Y}\right)
\right.\nonumber\\
&-&2\left.\partial_{[\mu}\left(\epsilon_{ABCDE} B^{AF}_{[\alpha} 
Y_F B^{BC}_{\beta ]} 
C_{\nu ]}^D\frac{T^{EG} Y_G}{Y\cdot T\cdot Y}\right)\right]
\label{mnab}\\
\frac{\sqrt{2}}{3m}
F_{\mu\alpha\beta\gamma}&=&\partial_\mu {\cal A}_{\alpha\beta\gamma}\nonumber\\
&-&\frac{4}{3}\partial_{\mu}\left(\epsilon_{ABCDE} B_{[\alpha}^{AB} 
B_\beta^{CF}Y_F 
B_{\gamma ]}^{DG}Y_G\frac{T^{EH} Y_H}{Y\cdot T\cdot Y}\right)\nonumber\\
&+&2\partial_{[\alpha}\left(\epsilon_{ABCDE} B_{\beta}^{AB} B_{\gamma ]}^{CF}
Y_F C_\mu^D \frac{T^{EG} Y_G}{Y\cdot T\cdot Y}\right)\nonumber\\
&+&\partial_\mu\left(\epsilon_{ABCDE} (\partial_{[\alpha}B_\beta^{AB} 
+\frac{4}{3} B_{[\alpha}^{AF} B_{\beta}^{FB})
B_{\gamma ]}^{CD} Y_E
\right)
\label{mabc}\\
\frac{\sqrt{2}}{3m}
F_{\alpha\beta\gamma\delta}&=&4\partial_{[\alpha} {\cal A}_{\beta\gamma
\delta ]}\nonumber\\
&+&4\partial_{[\alpha}\epsilon_{ABCDE}\left(-\frac{4}{3}B_\beta^{AB} 
B_{\gamma}^{CF} Y_F
B_{\delta ]}^{DG} Y_G \frac{T^{EH} Y_H}{Y\cdot T\cdot Y}\right.
\nonumber\\
&+&\left.(\partial_\beta B_\gamma^{AB} +\frac{4}{3} B_\alpha^{AF}
B_\gamma^{FB})
B_{\delta ]}^{CD} Y^E\right)
\label{abcd}
\eea
Here $T=T_{AB}\delta^{AB}$ and $Y\cdot T\cdot Y\equiv Y_AT^{AB}Y_B, Y\cdot T^2 
\cdot Y\equiv Y_A (T^2)^{AB}Y_B$.

This ansatz was obtained by requiring consistency of the susy laws, namely that
the the 11-dimensional susy variation law $\delta (d=11)F_{\Lambda\Pi\Sigma
\Omega}=d_{[\Lambda}\delta(d=11)A_{\Pi\Sigma\Omega}$ can be written as a total 
7-dimensional susy variation $\delta(d=7\; fields)$. Our present ansatz  
reproduces the linearized limit of ref. \cite{pnt}, and it coincides with
the geometrical proposal by \cite{hmm,fhmm} when we let $T^{AB}=\delta^{AB}$.
The $T$-dependent terms in (\ref{mnpr}) and (\ref{mnra}) 
which are $B$-independent separately satisfy the Bianchi identity, 
even though they are not an exact form.
The terms with ${\cal A}_{\alpha\beta\gamma}$ as well as the $B$ dependent 
terms are exact and thus they trivially satisfy the Bianchi identities.
The Chern-Simons terms in d=7 are not affected by the partial dressing with 
scalar fields of some spherical harmonics of the ansatz proposed by Freed et 
al. \cite{fhmm}.
The precise expression of the 4-form added in 
the $F_{\mu\nu\rho\sigma}$ sector is highly constrained. It must 
reproduce the linearized term in \cite{pnt}, and it must yield the correct 
scalar potential in d=7 after integrating over the compact space.
In order to perform this integral to which both the Einstein action and the 
kinetic action of the 3-index photon contribute, we start with the metric 
in the internal space and its inverse:
\be
g_{\mu\nu}=\Delta^{4/5} {C_\mu}^A {C_\nu}^B T^{-1}_{AB}
\;\;;\;\;
g^{\mu\nu}=\Delta^{2/5} \left( C^\mu_A C^\nu_B T^{AB} Y_C Y_D T^{CD}
-C^\mu_A Y_B T^{AB} C^\nu_C Y_D T^{CD}\right)
\ee
where $C_{\mu}^A=\partial_{\mu}Y^A$ is a conformal Killing vector.
We can thus interpret the deformations of the background metric as describing 
an ellipsoid with the conformal factor $\Delta^{4/5}$, 
whose axes at a specific point $y$ in the d=7 space time are determined by the 
eigenvalues of $T^{-1}_{AB}$. 
When setting the gauge fields to zero and disregarding the terms 
with d=7 space time derivatives, the integral over the 
compact space of the Einstein action
is already of the desired form, namely a linear 
combination of $T^2$ and $Tr(T^2)$.

On the other hand, the integrated kinetic action of the 3 index 
photon has  
the form $\int d^4 x \frac{-9}{4}\sqrt{det\sirc{g}(x)} 
(Y_E Y_F T^{EF})^2(1+{\cal S})^2$, 
where $(3/\sqrt{2})\sqrt{det\sirc{g}(x)} 
\epsilon_{\mu\nu\rho\sigma}{\cal S}$ is 
the extra term we need to add in $F_{\mu\nu\rho\sigma}$ besides its 
background value.
This function S should be of degree zero in $T$ and vanishes in the 
background. In order that the d=7 scalar potential be of the form 
$(Tr T)^2-2Tr T^2$, we can only admit terms in S of the form $\alpha
[Tr T/(Y\cdot T \cdot Y)-5]+\beta[Y\cdot T^2\cdot Y/(Y\cdot T\cdot Y)^2-1]$.
Requiring agreement with the linearized ansatz yields $\alpha=1/3$, while 
$\beta$ satisfies the quadratic equation $\beta(\beta+2/3)=0$ in order to 
reproduce the d=7 scalar potential. The solution $\beta=0$ does not produce 
the correct gravitino law, hence consistency requires $\beta=-2/3$.

The ansatz for the independent fluctuations ${\cal A}_{\alpha\beta\gamma}$ and 
the auxiliary field $E^{-1/2}B_{\alpha\beta\gamma\delta}$ is found
by matching the last term in $\delta\psi_{\alpha}$ in (II.6) (the 
term with $S_{\alpha\beta\gamma ,A}$):
\bea
\frac{i\sqrt{3}}{2}
{\cal A}_{\alpha\beta\gamma }&=&S_{\alpha\beta\gamma ,A}Y^A\label{indep}\\
\frac{i\sqrt{3}}{2}
\frac{B_{\alpha\beta\gamma\delta}}{\sqrt{E}}&=&[\frac{24k}{5}\bigtriangledown
_{[\alpha}S_{\beta\gamma\delta ],A}-\frac{k}{5}\delta_{AC}
{{\Pi^{-1}}_i}^C{{\Pi^{-1}}_j}^B
\delta^{ij}{\epsilon_{\alpha\beta\gamma\delta}}^{\epsilon\eta\zeta}S_{\epsilon
\eta\zeta ,B}]Y^A\label{auxil}
\eea
where the first terms in $B_{\alpha\beta\gamma\delta}$ cancel possible 
$\partial_{\alpha}S_{\beta\gamma\delta ,A}$ and $B_{\alpha}^{AB}S_{\beta\gamma
\delta ,B}$ terms in $\delta\psi_{\epsilon}$. We also find that $ka=-\frac
{5\sqrt{2}}{9}$ and $kb=-\frac{5\sqrt{2}}{72}$, fixing the free constants 
$a$ and $b$. However, since $B_{\alpha\beta\gamma\delta}=0$ has to be an 
equation of motion we should add to (\ref{auxil}) fermion bilinear terms 
and an FF term to complete the $S_{\alpha\beta\gamma ,A}$ equation of 
motion:
\be
\frac{i\sqrt{3}}{2}
\frac{B_{\alpha\beta\gamma\delta}}{\sqrt{E}}=-\frac{k}{5}
{\epsilon_{\alpha\beta\gamma\delta}}^{\epsilon\eta\zeta}\frac{\delta {\cal L}
^{(7d)}}{\delta S^{\epsilon\eta\zeta, A}}Y^A
\ee

At this moment all the ans\"{a}tze are fixed, and we can verify the remaining 
terms in the 7d susy transformation rules (II.4,II.6,II.7). 
This provides a number of independent nontrivial checks on all our 
ans\"{a}tze. The calculations involved in these checks will be publised 
elsewhere \cite{wip}. They involve heavy use of the formalism of spherical 
harmonics \cite{pvn}.

Finally, let us comment on applications to the AdS-CFT correspondence. The 
fact that there exists a consistent truncation means that we can use the
7 dimensional gauged sugra action for calculations of correlators of 
the operators in the 6d (0,2) CFT which correspond to the gauged supergravity
fields, at leading order in N. Indeed, consistency of the truncation means 
that there are no linear couplings of 'massive' fields to the gauged 
sugra, and so in the tree diagrams of 
gauged sugra the massive fields
will not appear. In \cite{cfm}, a computation of correlators
of chiral primary operators in the CFT was performed, following the work for 
the $AdS_5\times S_5$ case in \cite{lmrs}  (for other calculations of 3- and 
4-point functions see \cite{3point,4point}) . To find the 
correct CFT behaviour, a nonlinear redefinition of the scalar fields was 
also needed, which did result in a 
consistent truncation of the scalar modes to the massless ones. 
The nonlinear redefinition in $d=7$ is equivalent to our nonlinear 
embedding in $d=11$ for the massless modes, but note that the results of 
\cite {cfm} are only up to quadratic order (and only for the scalars)
whereas we do 
find a fully consistent truncation to all massless modes. With our results 
one can extend the calculations of CFT correlators to the other massless
(sugra) bosonic sectors and to the fermionic sector.

We expect that we can also find a consistent truncation in the $AdS_5
\times S_5$ case, in which  the same comments apply to the 
correspondence between $AdS_5\times S_5$ and
N=4 d=4 SYM. (Again, a consistent truncation of the scalar 
modes to the massless ones was implicitly obtained in \cite{lmrs}, by imposing
 the correct CFT behaviour). Perhaps our methods can also be used to complete 
the explicit expression for the truncation on $AdS_4\times S_7$.

{\bf Acknowledgements} We would like to thank B. de Wit for a useful discussion
on his work and I. Park for collaboration at the early stages of this work.
\nopagebreak

\end{document}